%% file: template.tex
\documentclass[master]{ws-rv9x6} % 'master' to combine all chapters
\usepackage[pdfpagelabels=false]{hyperref}  % \label, \ref and \cite are recommended
\usepackage{ws-rv-har}    % author-date citation/references %%% WS 04-Dec-17
\usepackage{ws-rv-thm}    % comment this line when `amsthm / theorem / ntheorem` package is used
\usepackage{subfigure}    % required only when side-by-side / subfigures are used
\usepackage{ws-index}     % required only when multiple indexes are used
\usepackage{enumitem}
\usepackage[percent]{overpic}
\usepackage{xcolor}
\usepackage{url}

\makeindex
\newindex{aindx}{adx}{and}{Author Index}       % author index
\renewindex{default}{idx}{ind}{Subject Index}  % subject index

\begin{document}		
\newcommand{\ltsima}{$\; \buildrel < \over \sim \;$}
\newcommand{\lsim}{\lower.5ex\hbox{\ltsima}}
\newcommand{\gtsima}{$\; \buildrel > \over \sim \;$}
\newcommand{\gsim}{\lower.5ex\hbox{\gtsima}}
\newcommand{\bra}{\langle}
\newcommand{\ket}{\rangle}
\newcommand{\lprime}{\ell^\prime}
\newcommand{\lpp}{\ell^{\prime\prime}}
\newcommand{\mprime}{m^\prime}
\newcommand{\mpp}{m^{\prime\prime}}
\newcommand{\ci}{\mathrm{i}}
\newcommand{\dd}{\mathrm{d}}
\newcommand{\veck}{\mathbf{k}}
\newcommand{\vecx}{\mathbf{x}}
\newcommand{\vecr}{\mathbf{r}}
\newcommand{\vecv}{\mathbf{\upsilon}}
\newcommand{\vecw}{\mathbf{\omega}}
\newcommand{\vecj}{\mathbf{j}}
\newcommand{\vecq}{\mathbf{q}}
\newcommand{\vecl}{\mathbf{l}}
\newcommand{\vecn}{\mathbf{n}}
\newcommand{\lm}{\ell m}
\newcommand{\that}{\hat{\theta}}
\newcommand{\thatp}{\that^\prime}
\newcommand{\chip}{\chi^\prime}
\newcommand{\hs}{\hspace{1mm}}
\newcommand{\nar}{New Astronomy Reviews}
\def\gsim{~\rlap{$>$}{\lower 1.0ex\hbox{$\sim$}}}
\def\lsim{~\rlap{$<$}{\lower 1.0ex\hbox{$\sim$}}}
\def\Msun {\,\mathrm{M}_\odot}
\def\Jcrit {J_\mathrm{crit}}
\newcommand{\rsun}{R_{\odot}}
\newcommand{\mbh}{M_{\rm BH}}
\newcommand{\Msunyr}{M_\odot~{\rm yr}^{-1}}
\newcommand{\mdot}{\dot{M}_*}
\newcommand{\ledd}{L_{\rm Edd}}
\newcommand{\cmc}{{\rm cm}^{-3}}
\def\gsim{~\rlap{$>$}{\lower 1.0ex\hbox{$\sim$}}}
\def\lsim{~\rlap{$<$}{\lower 1.0ex\hbox{$\sim$}}}
\def\Msun {\,\mathrm{M}_\odot}
\def\Jcrit {J_\mathrm{crit}}
\renewcommand{\thefootnote}{\arabic{footnote}}

\def\simgreat{\lower2pt\hbox{$\buildrel {\scriptstyle >}
   \over {\scriptstyle\sim}$}}
\def\simless{\lower2pt\hbox{$\buildrel {\scriptstyle <}
   \over {\scriptstyle\sim}$}}
\def\msobh{M_\bullet^{\rm sBH}}
\def\zodot{\,{\rm Z}_\odot}
\newcommand{\lambdabar}{\mbox{\makebox[-0.5ex][l]{$\lambda$} \raisebox{0.7ex}[0pt][0pt]{--}}}

\def\na{NewA}%
          % New~Astronomy
\def\aj{AJ}%
          % Astronomical Journal
\def\araa{ARA\&A}%
          % Annual Review of Astron and Astrophys
\def\apj{ApJ}%
          % Astrophysical Journal
\def\apjl{ApJ}%
          % Astrophysical Journal, Letters
\def\jcap{JCAP}

\def\pasa{PASA}

\def\apjs{ApJS}%
          % Astrophysical Journal, Supplement
\def\ao{Appl.~Opt.}%
          % Applied Optics
\def\apss{Ap\&SS}%
          % Astrophysics and Space Science
\def\aap{A\&A}%
          % Astronomy and Astrophysics
\def\aapr{A\&A~Rev.}%
          % Astronomy and Astrophysics Reviews
\def\aaps{A\&AS}%
          % Astronomy and Astrophysics, Supplement
\def\azh{AZh}%
          % Astronomicheskii Zhurnal
\def\baas{BAAS}%
          % Bulletin of the AAS
\def\jrasc{JRASC}%
          % Journal of the RAS of Canada
\def\memras{MmRAS}%
          % Memoirs of the RAS
\def\mnras{MNRAS}%
          % Monthly Notices of the RAS
\def\pra{Phys.~Rev.~A}%
          % Physical Review A: General Physics
\def\prb{Phys.~Rev.~B}%
          % Physical Review B: Solid State
\def\prc{Phys.~Rev.~C}%
          % Physical Review C
\def\prd{Phys.~Rev.~D}%
          % Physical Review D
\def\pre{Phys.~Rev.~E}%
          % Physical Review E
\def\prl{Phys.~Rev.~Lett.}%
\def\pasp{PASP}%
          % Publications of the ASP
\def\pasj{PASJ}%
          % Publications of the ASJ
\def\qjras{QJRAS}%
          % Quarterly Journal of the RAS
\def\skytel{S\&T}%
          % Sky and Telescope
\def\solphys{Sol.~Phys.}%
          % Solar Physics

          % Solar Physics
\def\sovast{Soviet~Ast.}%
          % Soviet Astronomy
\def\ssr{Space~Sci.~Rev.}%
          % Space Science Reviews
\def\zap{ZAp}%
          % Zeitschrift fuer Astrophysik
\def\nat{Nature}%
          % Nature
\def\iaucirc{IAU~Circ.}%
          % IAU Cirulars
\def\aplett{Astrophys.~Lett.}%
          % Astrophysics Letters
\def\apspr{Astrophys.~Space~Phys.~Res.}%
          % Astrophysics Space Physics Research
\def\bain{Bull.~Astron.~Inst.~Netherlands}%
          % Bulletin Astronomical Institute of the Netherlands
\def\fcp{Fund.~Cosmic~Phys.}%
          % Fundamental Cosmic Physics
\def\gca{Geochim.~Cosmochim.~Acta}%
          % Geochimica Cosmochimica Acta
\def\grl{Geophys.~Res.~Lett.}%
          % Geophysics Research Letters
\def\jcp{J.~Chem.~Phys.}%
          % Journal of Chemical Physics
\def\jgr{J.~Geophys.~Res.}%
          % Journal of Geophysics Research
\def\jqsrt{J.~Quant.~Spec.~Radiat.~Transf.}%
          % Journal of Quantitiative Spectroscopy and Radiative Trasfer
\def\memsai{Mem.~Soc.~Astron.~Italiana}%
          % Mem. Societa Astronomica Italiana
\def\nphysa{Nucl.~Phys.~A}%

\def\physrep{Phys.~Rep.}%
          % Physics Reports
\def\physscr{Phys.~Scr}%
          % Physica Scripta
\def\planss{Planet.~Space~Sci.}%
          % Planetary Space Science
\def\procspie{Proc.~SPIE}%
          % Proceedings of the SPIE

\newcommand{\rmp}{Rev. Mod. Phys.}
\newcommand{\ijmpd}{Int. J. Mod. Phys. D}
\newcommand{\sovjetp}{Soviet J. Exp. Theor. Phys.}
\newcommand{\jkas}{J. Korean. Ast. Soc.}
\newcommand{\PPVI}{Protostars and Planets VI}
\newcommand{\njp}{New J. Phys.}
\newcommand{\rap}{Res. Astro. Astrophys.}

\input{ch9.tex}

{
\bibliographystyle{ws-rv-har}    % author-date citation/references %%% WS 04-Dec-17
\bibliography{ref}
}

\printindex[aindx]           % to print author index
\printindex                  % to print subject index

\end{document}

%% file: ch9.tex
\setcounter{chapter}{8}
\title{Formation of the First Black Holes}
 \chapter[Statistical predictions for the first black holes]{Statistical predictions for the first black holes$^1$}%\footnotemark[1]}
 \label{chap:BHstat}

\author[Tilman Hartwig]{Tilman Hartwig}

\address{The University of Tokyo,\\{Kavli IPMU (WPI), UTIAS, \\ Kashiwa, Chiba 277-8583, Japan, \\ Tilman.Hartwig@ipmu.jp}}

\begin{abstract}
The recent observations of supermassive black holes (SMBHs) at high redshift challenge our understanding of their formation and growth. There are different proposed pathways to form black hole (BH) seeds, such as the remnants of the first stars (chapter~4), gas-dynamical processes (chapter~5), direct collapse (chapter~6), or stellar collisions in dense nuclear clusters (chapter~7). In this chapter, we discuss the probability of forming supermassive black holes (SMBHs) via these channels and determine the expected number density of the BH seeds. We start with a brief discussion of the observational constraints on SMBHs at low and high redshift that theoretical models have to reproduce (a more detailed account is provided in chapter~12). We further present the most popular formation channels of SMBHs, discuss under which conditions they can reproduce the observations, and compare various estimates in the literature on the expected number density of SMBHs. To account for the density of quasars at $z>6$ requires very efficient gas accretion mechanisms or high BH seeds masses. The bottleneck to obtain sufficiently high number densities of seed BHs with masses $>10^5$~$\Msun$ is the interplay between radiative and chemical feedback, which constrains the conditions for primordial, isothermal gas collapse.
\end{abstract}
%\markright{Customized Running Head for Odd Page} % default is Chapter Title.
\body

\setcounter{page}{161}
\footnotetext{$^1$ Preprint~of~a~review volume chapter to be published in Latif, M., \& Schleicher, D.R.G., ``Statistical predictions for the first black holes", Formation of the First Black Holes, 2018 \textcopyright Copyright World Scientific Publishing Company, \url{https://www.worldscientific.com/worldscibooks/10.1142/10652}}

\section{Observational Constraints}%\index{floats!figures}
Although BHs do not emit electromagnetic radiation, the accretion disk around supermassive black holes (SMBHs) converts gravitational energy into heat and the emission of the brightest of these active galactic nuclei can be seen as quasars up to high redshift. From the optically selected quasar sample in the SDSS survey we can determine a lower limit for the abundance of SMBHs with $>10^9\Msun$ at $z=6$ of about 1/\,cGpc$^{3}$  \citep{Fan2006,venemens13,deRosa14}. As we will see below, current models predict sufficiently many massive BH seeds to explain the number of quasars at $z>6$. However, the direct collapse (DC) scenario, which produces massive BH seeds (see chapter~5), cannot account for the population of less massive BHs at lower redshift under conservative parameter assumptions \citep{Greene2012,reines13}. Moreover, not all halos are able to fuel gas to their centre at high rates over several hundred million years to grow the BH seeds to their final masses (see chapter~10 for the feedback limited growth rates).

The discovery of the Lyman-$\alpha$ emitter CR7 at $z=6.6$ \citep{Sobral} might yield another constraint on the number density of direct collapse black holes (DCBHs). The strong Lyman-$\alpha$ and He{\sc ii} emission with large equivalent widths, the steep UV slope, and the absence of metal lines in the original observation by \citet{Sobral} suggests a massive population of metal-free stars with a total stellar mass of $\sim 10^7\Msun$. However, several studies demonstrate that the observational signature is more likely to emerge from an accreting DCBH in the centre of CR7 \citep{Pallottini2015,Agarwal2016,Hartwig16,Smidt2016,Smith2016,Pacucci17} than by a population of metal-free stars, but see \citet{Visbal16,Bowler17}. The expected number density of CR7-like sources is of the order $10^{-6}-10^{-7}\,\mathrm{Mpc}^{-3}$ \citep{Pallottini2015,Hartwig16,Visbal16}, which yields an additional observational constraint on the abundance of the first BHs. However, see also recent observation that challenge the interpretation of CR7 as a DCBH \citep{Matthee17, Sobral17, Bowler17}.

\begin{table}[htp]
\caption{default}
\begin{center}
\begin{tabular}{|c|c|}
Abbreviation & full name\\ \hline
SMBH & supermassive black hole\\
BH & black hole\\
DC & direct collapse\\
DCBH & direct collapse black hole\\
SN & supernova\\
LW & Lyman-Werner\\
PDF & probability distribution function
\end{tabular}
\end{center}
\label{abrsuper}
\end{table}%

In the next section we will see that stellar mass BHs, as the remnants of the first stars (see chapter~4), are sufficiently abundant and it is not their number density that poses a problem, but rather the necessary mass accretion over more than six orders of magnitude and the associated feedback effects that pose the biggest challenge (see chapter~10). We will hence mostly focus on DCBHs, which are predicted to be born with a higher seed mass, but which form only under very peculiar conditions and it is worth to investigate the probability of these environmental conditions of DC to assess the likelihood and the halo occupation fraction for DCBHs.

\section{Probability for stellar mass seed black holes}
\begin{figure}
\centerline{\includegraphics[height=6cm]{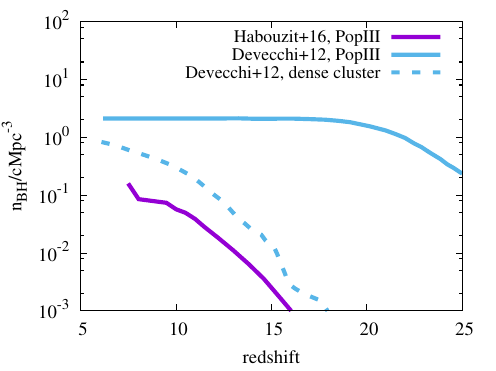}}
\caption{Comparison of the comoving number density of stellar mass BHs as a function of redshift based on models by \citet{Devecchi2012} and \citet{Habouzit2016a}. To convert from the given mass density in \citet{Habouzit2016a} to number density we assume an average mass of $100\Msun$ for Pop~III remnant BHs and of $700\Msun$ for BHs formed by runaway collisions in dense stellar clusters. The number of Pop~III seed black holes in \citet{Habouzit2016a} is a lower limit, because their simulation does not resolve minihalos as potential formation sites of these BHs. The number density of DCBHs is always below $10^{-5}\,\mathrm{cMpc}^{-3}$ \citep{Habouzit2016a}, but see chapter~5 for a more detailed discussion.}
\label{fig:ncomp1}
\end{figure}

Stellar mass seed BHs of $100-1000\Msun$ can either form as remnants of the first metal-free stars (see chapter~4) or as a consequence of runaway stellar collisions in metal poor nuclear clusters (see chapter~7). \citet{Devecchi2012} estimate the seed density for these two scenarios with a semi-analytical model of structure formation. Their model confirms recent observations that all halos with masses above $10^{11}\Msun$ host a BH and predicts a BH occupation fraction of $10\%$ for halos less massive than $10^9\Msun$. \citet{Habouzit2016a} demonstrate that the number density of Pop~III remnant BHs is $\gtrsim 10^{-2}\,\mathrm{cMpc}^{-3}$ at $z<10$, significantly higher than the number density of DCBH seeds in their numerical simulations. We compare the predicted number densities of stellar mass BHs in Fig. \ref{fig:ncomp1}, see also Fig 4, chapter~7.
The number density of BH seeds formed by runaway stellar collisions in nuclear clusters in slightly below the number density of Pop~III remnant BHs, but both are significantly above the expected number density of DCBHs (see section \ref{sec:pDC}). However, since BH formation via runaway collisions in dense stellar clusters can also occur in slightly metal-enriched gas, this formation channel dominates the mass density of BH seeds at $z<10$ \citep{Devecchi2012}. In their model, BH form via runaway stellar collision only in the narrow metallicity range $10^{-3.5}<Z/Z_\odot<10^{-3}$, but recent simulations show a wider range of conditions for this BH formation channel \citep{Sakurai2017, Reinoso18, Boekholt18}.

Hence, the bottleneck to form SMBHs at high redshift from stellar mass BHs is not the abundance of their seeds but rather the required high accretion rate over a sufficiently long time and the associated feedback (see chapter~10). In the following section we will hence focus on DCBHs and quantify the probability to form such a massive BH seed.

\section{Probability for the direct collapse scenario}\label{sec:pDC}
The idea of forming a massive BH as the result of the monolithic collapse of a proto-galactic gas cloud was first proposed by \citet{Rees1984}. Since then, various flavours of this formation scenario have emerged with different processes triggering the collapse of the cloud: galaxy mergers \citep{Mayer2010}, gas instabilities \citep{Begelman2006,Choi2013}, or contraction due to radiative cooling \citep{Bromm03}.

The main requirement for these scenarios is a high mass infall rate so that the accretion timescale of the infalling gas is shorter than the Kelvin-Helmholtz timescale of the protostar \citep{Hosokawa2013,Schleicher13,Sakurai2016}. Under this condition, the stellar radius monotonically increases with mass and enough gas can be accreted before the star reaches the main sequence of hydrogen burning. In this section we mostly focus on the DC scenario as a massive BH seed forming as the result of the thermodynamically triggered collapse of a pristine gas cloud with high mass infall rates and suppressed fragmentation. We will also comment on alternatives and explore the formation, such as the formation of massive seed BHs in metal poor gas or under the influence of dust cooling.

Pristine halos that cool mainly by atomic hydrogen are expected to have high mass infall rates, because they isothermally collapse at $T\approx 10^4$\,K and the inefficient atomic cooling prevents fragmentation of the gas and allows for high mass infall rates ($\dot{M}_\mathrm{in} \propto T^{3/2}$). A photodissociating flux from a nearby star-forming galaxy is required to keep the gas atomic. Phrased differently, a pair of halos has to be well synchronised in space and time to trigger DC \citep{Visbal2014,Regan17}: they have to be close enough to generate a sufficiently high photodissociating flux, but they have to be sufficiently far separated to prevent metal enrichment and tidal stripping \citep{chon16,Regan17}. Regarding the temporal synchronisation, there is only a short window after the initial starburst in which the LW flux is high enough, but the metal-enriched, supernova (SN)-driven winds have not yet reached the neighbouring halo \citep{Dijksta2014,Habouzit2016a,Hartwig16,Agarwal2016}.

\subsection{Atomic cooling halos}
The cooling function of hydrogen rises steeply around $\sim 8000$\,K (see Fig. 3 in chapter~3) and we follow the convention to define halos with $T_\mathrm{vir} \gtrsim 10^4$\,K as atomic cooling halos. The virial temperature of a halo of mass $M$ at redshift $z$ is given by
\begin{equation}
T_\mathrm{vir}=2\times 10^4 \left( \frac{M}{10^8\Msun} \right) ^{2/3} \left( \frac{1+z}{10} \right) \,\mathrm{K}
\end{equation}
for atomic gas of primordial composition \citep{Loeb2010} and the critical mass for an atomic cooling halo as a function of redshift is hence given by
\begin{equation}
M_\mathrm{ac} = 3.5 \times 10^7\Msun \left( \frac{1+z}{10} \right)^{-3/2}.
\label{eq:Mac}
\end{equation}
The number density of atomic cooling halos can be calculated via Press-Schechter theory \citep{Press1974} as outlined in chapter~2. The latter is based on the assumption of a Gaussian random fluctuation field, employing the results of linear theory to predict when the evolution becomes non-linear. The approach adopted here includes the corrections by \citet{smt01}, yielding:
\begin{equation}
n_\mathrm{ac} = \int _{M_\mathrm{ac}} ^{\infty} \sqrt{\frac{2}{\pi}} \frac{\rho _m}{M} \frac{-\mathrm{d}(\ln \sigma)}{\mathrm{d}M} \nu _c \exp (- \nu _c ^2/2)\,\mathrm{d}M,
\end{equation}
with $\nu _c = \delta _\mathrm{crit}(z)/\sigma (M)$, where $\delta _\mathrm{crit}(z)$ is the critical overdensity for collapse, $\sigma (M)$ is the variance of the density power spectrum at mass $M$, and $\rho _m$ is the mean matter density of the Universe. However, this constrain is not very restrictive because any host galaxy of a SMBH once passed this mass threshold of atomic cooling and was hence a potential candidate for DC. In the next section we will discuss more constraining criteria such as the photodissociating radiation and the absence of metals.

\subsection{Pristine Gas}
In order to prevent fragmentation and maintain a high temperature and hence mass infall rate of the collapsing gas cloud, the halo has to be pristine (but see \citet{Omukai2008,Latif2016D,Agarwal17} for alternative scenarios). Metals are a much more efficient coolant than hydrogen and the higher cooling rate makes gas more susceptible to fragmentation and prevents DC via ordinary star formation. Halos can either be enriched internally by Pop~III stars or externally by SN-driven metal-enriched winds from nearby star-forming halos and external enrichment seems to be more efficient in suppressing the formation of DCBHs \citep{Dijksta2014}.
See Figure 2 which illustrates how the fraction of pristine gas and halos evolves with redshift.

\begin{figure}
\vspace{0.3cm}
\centerline{\includegraphics[height=8cm,origin=c]{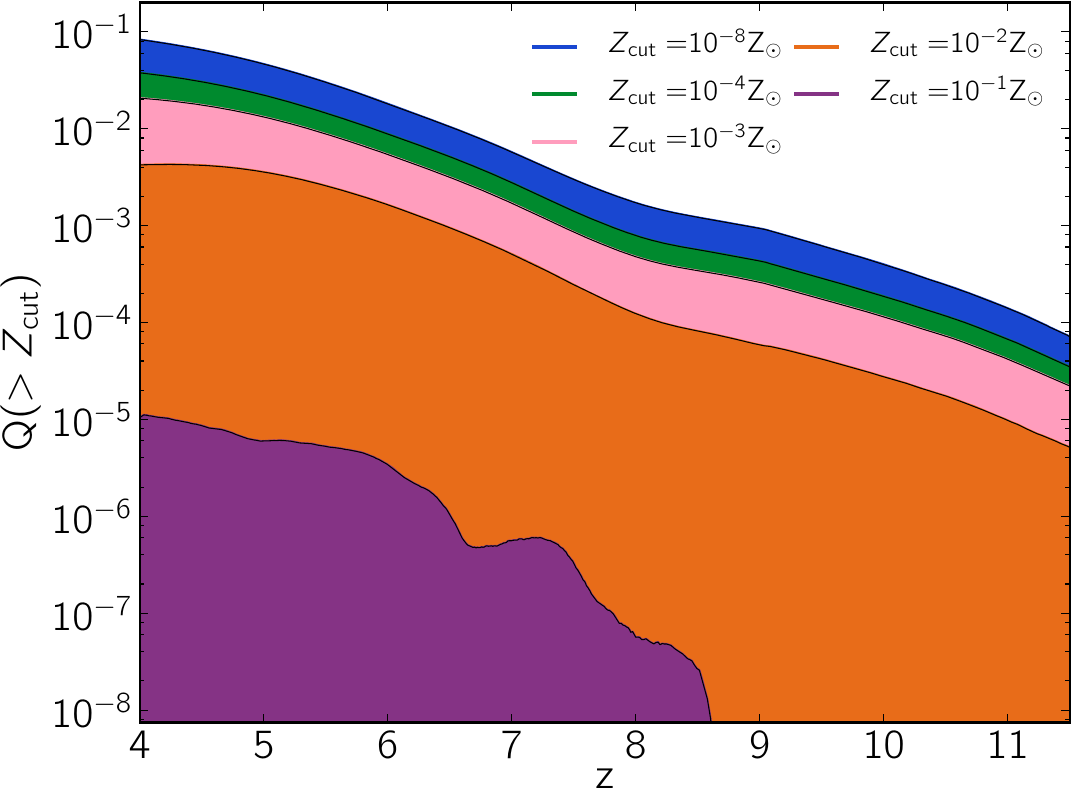}}
\caption{Left: Redshift evolution of the metal volume filling factor for different values of the metallicity cut, $Z_\mathrm{cut}$. Figure adopted from \citet{Pallottini2014}, reproduced by permission of Oxford University Press / on behalf of the RAS.}
\label{fig:zZ}
\end{figure}
At $z=4$ only $\lesssim 10\%$ of the cosmological volume is enriched with metals, see also \citet{Johnson2013}. However, this volume filling factor also accounts for voids without star formation. When we consider atomic cooling halos as the halos of interest for DC, \citet{Latif2016D} find that more than half of them are still pristine until $z \approx 9$ (see Figure 1 of chapter~5).  These candidate halos have not been externally enriched and also star formation in these halos was suppressed by e.g. a photodissociating background \citep{machacek01}. Considering this criterion for DC independently may not seem very constraining, because every second target halo is still pristine at the redshift of interest. However, we discuss in the next section that DC generally requires a nearby star-forming galaxy, which drastically reduces the probability of not being polluted by metals.

\subsection{Photodissociating Radiation}
A photodissociating LW flux is another crucial ingredient for DC. The concept of a critical flux threshold, $J_\mathrm{crit}$, above which a DCBH can form is well established in the literature, but there is no consensus on its value and most likely it is a distribution of critical values depending on e.g. the spectral shape of the stellar population \citep{Shang2010,Sugimura2014}, the anisotropy of the illuminating flux \citep{Regan2014B,Regan2016}, the treatment of H$_2$ self-shielding \citep{WolcottGreen2011,Hartwig2015}, the details of the chemistry model \citep{Glover2015a,Glover2015b}, and the star formation history \citep{Agarwal16}. We refer here also to the detailed discussions on this topic in chapter~3 on chemistry and chapter~5 on the determination of $J_{\rm crit}$ from dynamical simulations and on its dependence on the radiation spectrum. Although there is no universal value for $J_\mathrm{crit}$, it is convenient to quantify the importance of different effects by this value. Once we have chosen a threshold value for DC, we can quantify the probability to find a pristine atomic cooling halo that is exposed to a LW flux above this threshold. In this section we will review and summarise different models to address this question and to derive the number density of DCBH formation sites. We use the standard notation for the LW flux and express its intensity in units of $J_{21}=10^{-21}\mathrm{erg}\,\mathrm{s}^{-1}\,\mathrm{cm}^{-2}\,\mathrm{sr}^{-1}\,\mathrm{Hz}^{-1}$.

The build-up of the cosmological LW background is self-regulated \citep{johnson08}, because a high flux suppresses star formation and hence the production of further LW photons. The mean value in the redshift range $z=10-20$ is of the order a few times $10\,J_{21}$ \citep{Dijkstra08,ahn09,holzbauer12,McQuinn12,Agarwal12,Dijksta2014,chon16}, but possible formation sites for DC are hidden in the high-end tail of the $J_\mathrm{LW}$ distribution.

\citet{ahn09} present the first self-consistent radiative transfer calculations of the inhomogeneous LW background. By using high-resolution N-body simulations and an equivalent gray opacity for the radiative transfer they show that the LW background is inhomogeneous and correlates with the clustering of large-scale structure. In contrast to ionising photons in the high redshift Universe, LW photons can travel cosmological distances due to the small pre-reionisation H$_2$ abundance of $\sim 10^{-6}$ and hence an optical depth of $\tau _\mathrm{LW} < 1$ \citep{shapiro94}. However, due to the expansion of the universe they get redshifted out of the LW bands and can no longer contribute to H$_2$ photodissociation once their energy is below $\sim 11.5$\,eV (see also \citet{Haiman2000} for a discussion of radiative transfer of LW at high redshift on cosmological scales).

\citet{ahn09} define the ``LW horzion" as the maximum comoving distance from a source that an H$_2$ dissociating photon can reach and determine this radius to be $\sim 100\,\alpha$\,cMpc, with
\begin{equation}
\alpha = \left( \frac{h}{0.7} \right)^{-1} \left( \frac{\Omega _m}{0.27} \right)^{-1/2} \left( \frac{1+z_s}{21} \right)^{-1/2},
\end{equation}
where $z_s$ is the redshift of the source. Within this LW horizon the cosmic expansion reduces the effective LW flux by
\begin{equation}
f_\mathrm{mod}=
\begin{cases}
1.7 \exp [-(r_\mathrm{cMpc}/116.29 \alpha)^{0.68}] -0.7 \quad \mathrm{for} \quad r_\mathrm{cMpc}/ \alpha \leq 100\\
0 \quad \mathrm{for} \quad r_\mathrm{cMpc}/ \alpha > 100,
\end{cases}
\end{equation}
where $r_\mathrm{cMpc}$ is the distance from the source in comoving Mpc.

Due to the long mean free path of LW photons, many sources contribute to the local flux and \citet{Dijkstra08} find that $>99\%$ of all halos are exposed to a LW flux within a factor of two of the mean value. They calculate the LW background based on the clustering of halos and account for the non-linear Eulerian bias of the two-point correlation function \citep{iliev03}, which fits the correlation function derived in N-body simulations. Accounting for this non-linear bias is crucial since we are especially interested in close pairs of halos, which are the most promising candidates for DC \citep{Dijkstra08,Visbal2014,Regan17}. The differential probability to find a halo of mass $M$ at redshift $z$ in the distance $r$ to an atomic cooling halo is given by
\begin{equation}
\frac{\mathrm{d}^2 P_1 (M,r,z)}{\mathrm{d}M \, \mathrm{d}r} = 4 \pi r^2 (1+z)^3 [1+\xi (M_\mathrm{ac},M,r,z)] \frac{\mathrm{d}n}{\mathrm{d}M},
\end{equation}
where $\xi (M_\mathrm{ac},M,r,z)$ is the two-point correlation function, which yields the excess probability to find two halos of masses $M_\mathrm{ac}$ and $M$ at distance $r$, and ${\mathrm{d}n}/{\mathrm{d}M}$ denotes the Press-Schechter mass function \citep{Press1974,smt01}. \citet{Dijkstra08} model the LW luminosity of each halo by randomly sampling from a log-normal distribution of UV luminosities per unit star formation rate, $P_2/\mathrm{d}\log (L_\mathrm{LW})$. The probability distribution function (PDF) of a given atomic cooling halo to be exposed to a LW flux $J_\mathrm{LW}$ is hence given by
\begin{equation}
\frac{\mathrm{d}P}{\mathrm{d}\log (J_\mathrm{LW})} = \int _{M_\mathrm{min}} ^{\infty} \mathrm{d}M \int _{r_\mathrm{min}} ^{\infty} \mathrm{d}r \frac{\mathrm{d}^2 P_1}{\mathrm{d}M \, \mathrm{d}r} \frac{P_2}{\mathrm{d}\log (L_\mathrm{LW})}.
\label{eq:pdfD08}
\end{equation}
\citet{Dijkstra08,Dijksta2014,Inayoshi2015} assume $M_\mathrm{min}=M_\mathrm{ac}$, but the choice of the minimal radius $r_\mathrm{min}$ is debated, because its value is crucial: small values will populate the high-end tail of the PDF because the flux scales with the inverse square of the distance, but a too small value is physically prohibited to prevent external enrichment of the atomic cooling by the star forming halo and to prevent tidal stripping due to dynamical effects \citep{chon16,Regan17}.

We refer to \citet{Dijkstra08} for a discussion of how the choice of $r_\mathrm{min}$ (and other model parameters) affects the PDF of $J_\mathrm{LW}$. The high-end tail of the PDF is dominated by close pairs of halos and the illuminating flux, as seen by the atomic cooling halo, should hence be highly anisotropic \citep{Regan2014B, Regan15b}. Moreover, the relative velocity of this close pair of halos is relevant for the escape fraction of LW photons out of the star-forming galaxy, because photons might be shifted into the LW bands in the far-field \citep{Schauer2015,Schauer17}.

In \citet{Dijksta2014} they improve their previous model by explicitly accounting for pollution by metal-enriched winds from the star forming galaxy. They approximate the radius of the metal enriched region around a star-forming halo by
\begin{equation}
r_\mathrm{metals} = 0.3\,\mathrm{kpc} \left( \frac{M_*}{10^5\Msun} \right)^{1/5}\left( \frac{n}{1\,\mathrm{cm}^{-3}} \right)^{-1/5} \left( \frac{t}{\mathrm{Myr}} \right)^{2/5},
\end{equation}
where $M_*$ is the stellar mass and $n$ the density of the gas into which the SN-driven metal bubble expands. Based on this metal polluted radius, \citet{Dijksta2014} correct the PDF of $J_\mathrm{LW}$ (Eq. \ref{eq:pdfD08}) by multiplying it with $\Theta (r-r_\mathrm{metals})$ to exclude atomic cooling halos as potential candidates for DC if they are too close and already polluted. This approach was also used in \citet{Habouzit2016b,Hartwig16} to estimate the likelihood of finding DCBH candidates.

A comparison of the literature on the PDF of $J_\mathrm{LW}$ can be seen in Figure~4.
\begin{figure}
\includegraphics[width=9cm]{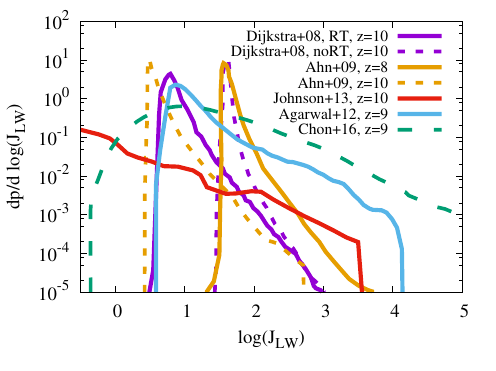}
\caption{Comparison of the PDF of the LW flux. The data is based on \citet{Dijkstra08,ahn09,Johnson2013,Agarwal12,chon16} (purple, orange, red, blue, and green, respectively). Note the slightly different redshifts and how the peak of this distribution shift towards higher fluxes with decreasing redshift due to the increasing cosmic star formation rate. All studies determine the LW flux based on the clustering of halos. \citet{chon16,Agarwal12} derive the distribution for pristine halos, \citet{ahn09} show the PDF for all grid cells inside their computational domain, \citet{Johnson2013} for primordial gas with $n \geq 1\,\mathrm{cm}^{-3}$, and \citet{Dijkstra08} provide the flux as seen by atomic cooling halos of mass $M=4 \times 10^7\Msun$. In their fiducial model (noRT) they assume the IGM to be transparent to LW photons and in their second model with radiative transfer they account for intergalactic H$_2$, which reduces the contribution from far away sources and hence the mean and minimum value of the PDF. However, the distribution at high $J_\mathrm{LW}$ is not affected by intergalactic absorption, because it is dominated by close pairs of halos.}
\label{fig:J21compilation}
\end{figure}
All studies broadly agree on the shape and mean value, which increases with decreasing redshift and all studies predict potential DCBH hosts, which are exposed to a flux above $J_\mathrm{crit}\approx 10^2-10^3$. The final density of DCBHs depends crucially on the high-end slope, which differs amongst the authors. While the recent simulation-based results of \citet{Agarwal12,chon16,Johnson2013} agree on the high-end slope, the models by \citet{Dijkstra08,ahn09} yield a steeper slope. \citet{Inayoshi2015}, who follow the approach by \citet{Dijkstra08,Dijksta2014}, find an even steeper slope of $\mathrm{d}p/\mathrm{d}\log (J_\mathrm{LW}) \propto J_\mathrm{LW}^{-5}$ in the range $10^3 \lesssim J_\mathrm{LW} \lesssim 10^4$. They attribute this discrepancy to the choice of $r_\mathrm{min}$.

The sharp cut-off at low $J_\mathrm{LW}$ is due to the very long mean free path of LW photons and hence the mean, median, and minimum value of this distribution are very close. The less sharp cut-off at low $J_\mathrm{LW}$ in \citet{Johnson2013,chon16} is caused by the smaller box size and the hence smaller number of sources contributing to the local flux. In the other studies, the box sizes are larger or they implicitly account for the LW background radiation by the cosmic star formation history.

Another crucial question is for how long an atomic cooling halo has to be exposed to a photodissociating flux above $J_\mathrm{crit}$ for it to collapse isothermally. The collapse time of the gas core of a typical atomic cooling halo is $t_\mathrm{coll} \approx 10$\,Myr at $z=10$ \citep{Visbal2014} with $t_\mathrm{coll} \propto (1+z)^{-3/2}$. Other authors require the conditions for DC to be fulfilled over at least the freefall time, which is of the order $100$\,Myr at the redshifts of interest and yields more conservative estimates \citep{Dijksta2014}. \citet{Habouzit2016a} analyse the impact of the choice of the collapse time on the probability to find DCBH candidates and find a significant difference: At $z=7.3$ they find 17 (3) DC formation sites in their computational volume for an assumed collapse time of 10\,Myr (150\,Myr).

\citet{chon16} demonstrate that the requirement of having a flux of at least $J_\mathrm{crit}$ at the time of virialisation of the atomic cooling halos is too strict for DC. They reduce the LW intensity artificially to $2\%$ of its original value at virialisation and still find the halo to collapse isothermally. They argue that the local flux increases over time due to the approaching halos by an order of magnitude and that lower intensities than previously thought are sufficient to induce DC, also see \citet{fernandez14}.

\subsection{Tidal and ram pressure stripping}
The conditions for DC are generally met in close halo pairs, where one halo provides the photodissociating radiation and the other one is still pristine and about to pass the mass threshold of atomic hydrogen cooling. However, dynamical effects such as tidal forces and ram pressure stripping can no longer be neglected for the thermal evolution of the gas collapse \citep{chon16}. Tidal stripping occurs in close pairs of galaxies, where the tidal force exerted by the more massive galaxy results in an effective transfer of gas and stars from the lower mass to the more massive galaxy. Ram pressure stripping can expel gas out of the gravitational potential of a galaxy that moves with a sufficiently high relative velocity with respect to the ambient medium.

To study these dynamical effect of the environment, \citet{chon16} select DC candidate halos in a semi-analytical model and perform 3D hydrodynamical follow-up simulations of these pristine atomic cooling halos that are illuminated by a sufficiently high LW flux. Out of 42 selected halos they find only two successful candidates and show that tidal forces and ram pressure stripping prevent the gas collapse in the other cases.

They also point out that while minor mergers tend to disrupt the gas distribution and prevent collapse, major mergers rather help in triggering the collapse due to the induced gravitational instability at the halo centre, caused by the asymmetric mass distribution on large scales, also see \citet{Mayer2010,Mayer2015}. In summary, they find that only $5\%$ of the atomic cooling halos that are pristine and illuminated with $J_\mathrm{LW}\geq J_\mathrm{crit}$ eventually collapse to form a DCBH.

\subsection{Density of direct collapse seed black holes}
\begin{figure}
\label{fig:ncomp}
\centerline{\includegraphics[width=\textwidth]{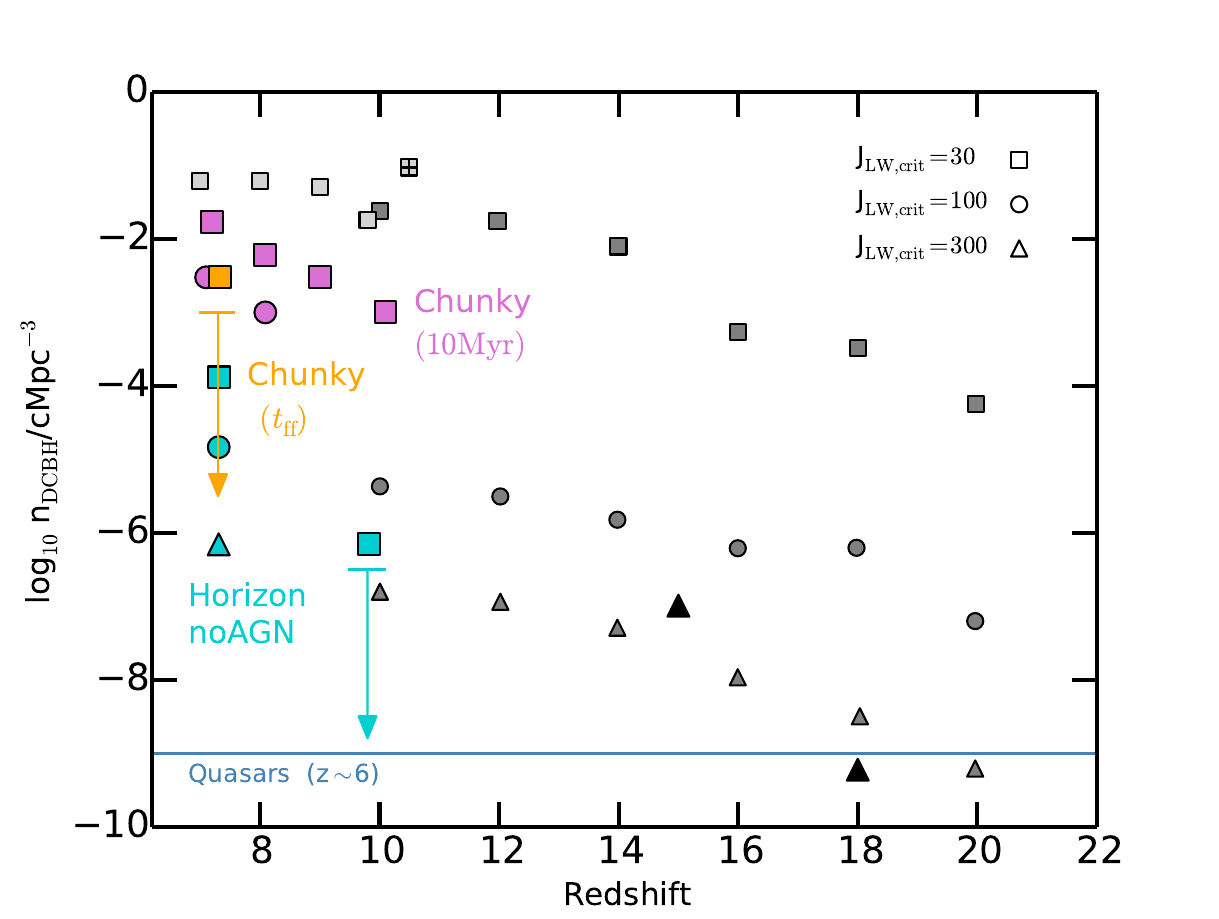}}
\caption{Comoving number density of DCBH formation sites, as a function of redshift. Symbol shapes represent different radiation intensity thresholds. Squares: $J_\mathrm{crit}=30$, circles: $J_\mathrm{crit}=100$, triangles: $J_\mathrm{crit}=300$. The horizontal solid blue line shows the comoving number density of quasars at $z\sim 6$. The light gray crossed square at $z = 10.5$ is from the hydrodynamical simulation by \citet{Agarwal14}, the light gray squares in the range $z = 7-10$ are from \citet{Agarwal12}, dark gray squares and black triangles are the results of \citet{Dijksta2014} and \citet{valiante16}, respectively. The orange square and purple symbols show the number density for \citet{Habouzit2016b} for different assumptions on the collapse time. The cyan squares, circle and triangle represent the large-scale cosmological simulation Horizon-noAGN (Dubois et al. 2014b). Adapted from \citet{Habouzit2016b}, reproduced by permission of Oxford University Press / on behalf of the RAS.}
\end{figure}

In this section, we compare different estimates for the probability to form DCBHs, their expected number density, and the resulting occupation fractions. As we have seen before, the PDF to find pristine atomic cooling halos exposed to a certain LW flux (Eq. \ref{eq:pdfD08}) can be calculated analytically \citep{Dijkstra08,Dijksta2014,Inayoshi2015}, by post-processing cosmological N-body simulations \citep{ahn09,Agarwal12}, or by performing cosmological simulations \citep{Johnson2013,Agarwal14,Habouzit2016b,chon16}. The number density of DCBHs can then be approximated by
\begin{equation}
n_\mathrm{DCBH}(z) = \int _{M_\mathrm{ac}} ^\infty \mathrm{d}M \frac{\mathrm{d}n}{\mathrm{d}M} P_\mathrm{DCBH}(\geq J_\mathrm{crit},z).
\end{equation}
However, this approach assumes that all pristine atomic cooling halos that are exposed to $>J_\mathrm{crit}$ collapse to a DCBH, but the analytical and semi-analytical models do not account for hydrodynamical effects such as tidal or ram pressure stripping. \citet{chon16} show that these effects can reduce the number number density of DCBHs by more than one order of magnitude. A comparison of the number density of DCBH as a function of redshift by different authors can be seen in Figure \ref{fig:ncomp}.

The number density increases with decreasing redshift and with decreasing $J_\mathrm{crit}$, but even for the same value of $J_\mathrm{crit}$ the absolute values of the number density differ by several orders of magnitude. All studies agree that chemical enrichment from previous star formation in progenitor halos or by metal-enriched winds from nearby star-forming halos play a crucial role in determining the final number density of DCBHs. Also the implementation of SN feedback strongly affects the possibility to form a DCBH \citep{Habouzit2016b}. Hydrodynamical simulations suffer from the limited volume which allows in some cases only for upper limits on the DCBH density \citep{Agarwal14,Habouzit2016b}. Other simulations find values of $2.5 \times 10^{-4}\,\mathrm{cMpc}^{-3}$ for $J_\mathrm{crit}=100$ \citep{chon16}
and  $(0.1-5)\times 10^{-6}\,\mathrm{cMpc}^{-3}$ for $J_\mathrm{crit}=300$ \citep{Habouzit2016a}. Using an analytical estimate for the clustering of halos, \citet{yue14} find that the DCBH mass density rises from $\sim 5\Msun\,\mathrm{Mpc}^{-3}$ at $z \sim 30$ to the peak value $\sim 5 \times 10^5 \Msun\,\mathrm{Mpc}^{-3}$ at $z\sim14$ in their fiducial model. However, the abundance of accreting DCBHs decreases after $z\sim 14$, because potential formation sites get either polluted or photoevaporated. We refer the interested reader to \citet{Habouzit2016b} for a more detailed quantitative comparison of the different semi-analytical models.

\citet{Bellovary11} use cosmological simulations to derive the BH occupation fraction and they show that halos with $M > 3\times 10^9\Msun$ host massive BHs regardless of the efficiency of seed formation. \citet{Habouzit2016b} find, based on post-processing state-of-the-art cosmological simulations \citep{Dubois14}, that $\sim 30\%$ of halos more massive than $10^{11}\Msun$ at $z=6$ have at least one progenitor that fulfils the conditions for DC, assuming a critical value of $J_\mathrm{crit}=30$. However, this occupation fraction is a strong function of the critical flux and falls below $1-10\%$ for $J_\mathrm{crit}>100$. In \citet{Habouzit2016a} they confirm an occupation of $\sim 10\%$ for Pop~III remnant BHs and an occupation fraction of the order $10^{-6}-10^{-5}$ for DCBHs, also see \citet{Tanaka2009ApJ}.

Although there are still large theoretical uncertainties in the number density of DCBH formation sites, caused by e.g. the model of metal enrichment, the value of $J_\mathrm{crit}$, or the necessary duration of the photodissociating feedback, all models are able to yield more than 1 DCBH per Gpc$^{3}$ by $z=6$, which is the lower limit set by current observations. The most optimistic scenarios are even able to explain the majority of BHs in local galaxies via the DC scenario. The remaining bottleneck is to grow these seed BHs to SMBHs by circumventing the self-regulating accreting feedback as we will see in the next chapter.

In this chapter, we presented the rather consensual picture on the statistical predictions on BH formation in the early Universe. However, we note that most presented results are subject of ongoing debate and reality might lie beyond what we currently assume to be realistic. We have presented DC as a promising pathway to form massive BH seeds, but other formation channels than the isothermal collapse may also be relevant, or different scenarios, such as runaway collisions in the first stellar clusters could be more efficient than we have assumed \citep{Katz2015,Sakurai2017}. Some predictions based on the latter scenario were therefore presented in chapter~7. Within the framework of the DC, the conservative conditions that the gas for the BH formation via gas-dynamical processes has to be pristine might be too strict and BH seed formation may also be possible in gas that is slightly enriched with metals or dust \citep{Omukai2008,Mayer2015,Latif2016D,Agarwal17}, see also discussion in chapter~6. All these caveats affect the predicted formation efficiency and number density of BH seeds in the early universe.

The statistical predictions on the abundances of seed black holes are of course only the first step. In the next chapter~10, we will describe their further growth in the presence of feedback, while scenarios for super-Eddington accretion are outlined in chapter~11. The subsequent chapters describe both the current observational status as well as future prospects.